\documentclass[superscriptaddress,preprintnumbers,aps,prl,showpacs,twocolumn]{revtex4}
\usepackage{epsf,epsfig,graphicx}
\usepackage{color}
\begin{document}

\preprint{MSUHEP-110720}

\title{QCD resummation for jet substructures}

\author{Hsiang-nan Li}
\email{hnli@phys.sinica.edu.tw}
\affiliation{Institute of Physics, Academia Sinica, Taipei, Taiwan 115, Republic of China,}
%\affiliation{Department of Physics, National Cheng-Kung University, Tainan, Taiwan 701, Republic of China}
%\affiliation{Department of Physics, National Tsing-Hua University, Hsin-Chu, Taiwan 300, Republic of China}
\author{Zhao Li}
\email{zhaoli@pa.msu.edu}
\affiliation{Department of Physics and Astronomy, Michigan State University, East Lansing, Michigan 48824, USA}
\author{C.-P. Yuan}
\email{yuan@pa.msu.edu}
\affiliation{Department of Physics and Astronomy, Michigan State University, East Lansing, Michigan 48824, USA}
\affiliation{Center for High Energy Physics, Peking University, Beijing 100871, China}


\begin{abstract}

We provide a novel development in jet physics by predicting the
energy profiles of light-quark and gluon jets in the framework of
perturbative QCD. Resumming large logarithmic contributions to all
orders in the coupling constant, our predictions are shown to agree
well with Tevatron CDF and Large-Hadron-Collider CMS data. We also
extend our resummation formalism to the invariant mass distributions
of light-quark and gluon jets produced in hadron collisions. The
predicted peak positions and heights in jet mass distributions are
consistent with CDF data within uncertainties induced by parton
distribution functions.

\end{abstract}

\maketitle

It has been a long-standing challenge to predict substructures
(including energy profiles and masses) of light-quark and gluon jets
in the perturbative QCD (pQCD) theory. During the Tevatron run 1 era
in the early 1990s, it was found that next-to-leading-order (NLO) QCD
calculations cannot describe experimental data on jet substructures.
Hence, it has been a custom for experimentalists to compare jet
substructures measured at the Tevatron, either at run 1 or run 2,
with predictions from full event generators such as PYTHIA or
HERWIG. While the full event generators (usually with specific tuning)
could describe data, it
remains desirable to develop a theoretical framework for the study
of jet substructures. 
For that, the soft-collinear effective theory was adopted in the literature, such as 
\cite{Ellis:2010rwa,Kelley:2011tj}.
In this Letter, we propose a novel
approach to predicting jet substructures based on the pQCD
resummation formalism\cite{Collins:1984kg}.
We show that results
of the resummation formalism
for light-quark and gluon  jets are well consistent with the
energy profiles measured by CDF at Tevatron  \cite{Acosta:2005ix} and CMS
at Large Hadron Collider (LHC) \cite{CMSJE}, and with the mass distributions measured by
CDF \cite{CDFJM}.

It is known that the top quark is predominantly produced at rest at the
Tevatron and can be clearly identified by detecting three (or more)
isolated jets from its hadronic decays. However, this strategy will
not work for identifying a highly boosted top quark
\cite{Agashe:2006hk,Fitzpatrick:2007qr,Baur:2007ck,Brooijmans:2008se}
at the LHC, which results in a single jet.
Furthermore, it has been pointed out
\cite{Skiba:2007fw,Holdom:2007ap} that an energetic QCD (light-quark
or gluon) jet can have  an invariant
mass around the top-quark mass to fake a top-quark jet.
That is, a boosted top-quark jet is
difficult to be discriminated against an ordinary QCD jet.
This difficulty also appears in the identification of a highly boosted
Higgs boson decaying into a bottom-quark pair
\cite{Butterworth:2008iy,Gabrielli:2007wf}, or a highly boosted $W$
or $Z$ boson decaying into hadronic final states, for they can all
produce a single-jet experimental signature. In order to improve
jet identification at the LHC, additional information from jet
energy profiles is needed, because jets initiated by different
parent particles usually produce different energy profiles.

We denote the jet energy function as $J^E_f(M_J^2,P_T,\nu^2,R,r)$ for defining a
light-quark ($f=q$) or gluon ($f=g$) jet with mass $M_J$, transverse
momentum $P_T$, and cone size $R$, which describes the all-order
energy distribution within a smaller cone of size $r<R$. $J^E_f$ is
constructed by inserting a sum of the step functions
$\sum_ik_{iT}\Theta(r-\theta_i)$ into the usual jet definition
\cite{Almeida:2008tp}, where $k_{iT}$ and $\theta_i$ are the
transverse momentum and the angle of the final-state particle $i$
with respect to the jet axis. At leading-order (LO), it is
a $\delta$-function, i.e.,  $J^{E(0)}_f=P_T\delta(M_J^2)$, which is
independent of $r$, because $\theta=0$ at this order. The jet
definition contains a Wilson line along the light cone, which
collects gluons emitted from other parts of the collision process and
collimated to the parent particle of the jet.
To employ the resummation
technique, we vary the Wilson line into an arbitrary direction
$n^\mu$ with $n^2\neq 0$ \cite{Li:1995eh}.
The dependence of $J^E_f$  on
$n^\mu$ and the jet momentum $P_J$ appears through the invariants
$n^2$, $P_J^2=M_J^2$, and $P_J\cdot n$ (which is related to $P_T$).
When $r$ approaches zero, the phase space of real radiation is
strongly constrained, so the infrared enhancement in real radiation
does not cancel completely with that in virtual correction. The resultant
large logarithms of the ratio $(P_J\cdot n)^2/(n^2r^2)$, which is
conveniently defined as $[R^2P_T^2/(4r^2)]\nu^2$,
should be resummed to all orders in the
coupling constant $\alpha_s$. It is easy to see from the above ratio
that the variation in $n$, i.e., $\nu^2$, can turn into the variation
in $r$.
To compare with present experimental data on jet energy profile , we consider the
jet energy function with the jet invariant mass being integrated out, which
corresponds to taking the $N=1$ moment in the Mellin space and is denoted as
${\bar J}_f^E(1,P_T,\nu^2,R,r)$.
By definition, different choices of $n^\mu$ yield the same
collinear divergences associated with the jet. Hence,
the effect of varying $n$, i.e., $\nu^2$, does not involve the
collinear divergences and can be factorized out of the jet energy function,
leading to an evolution equation
\begin{eqnarray}
\nu^2\frac{d}{d\nu^2}{\bar J}_f^E
%(1,P_T,\nu^2,R,r)
=\left[G^{(1)}+K^{(1)}\right]
%\nonumber\\
%& &\times
{\bar J}_f^E.
%(1,P_T,\nu^2,R,r).
\end{eqnarray}
The one-loop
kernel $G^{(1)}$ ($K^{(1)}$) absorbs the hard (soft) dynamics of the
variational effect, whose expressions are similar to those derived
in \cite{Li:1996gi}, but with the step function in angle being
inserted into the real-gluon piece of $K^{(1)}$.

The solution describing the evolution of the jet energy function
from the initial value $\nu^2_{\rm in}=C_1^2r^2/(C_2^2R^2)$ to the final value
$\nu^2_{\rm fi}=1$ is written as
\begin{eqnarray}
&& {\bar J}_f^E(1,P_T,\nu^2_{\rm fi},R,r)= {\bar
J}_{f}^E(1,P_T,\nu^2_{\rm in},R,r)
\nonumber \\ && \times
\exp\left\{
-\int^{C}_{C \nu_{\rm in}^2}\frac{dy}{y}
%\right. \nonumber \\ && \times \left.
\left[\frac{1}{2} \int_{y \nu_{\rm in}^2}^{y^2}\frac{d\omega}{\omega}
%\right.\right. \nonumber \\ && \times \left.\left.
A(\alpha_s(\omega C_2^2 R^2 P_T^2))
\right.\right. \nonumber \\ && \left.\left.
-\frac{C_f}{\pi} \alpha_s\left(y^2C_2^2 R^2 P_T^2\right)
\left(\frac{1}{2}+\ln\frac{C_2}{C_1}\right)
\right]\right\}, \label{ep1}
\end{eqnarray}
with the cusp anomalous dimension %$A=\sum_i(\alpha_s/\pi)^iA^{(i)}$, where
\begin{eqnarray}
A=\frac{\alpha_s}{\pi}C_f
+\frac{\alpha_s^2}{\pi^2}C_f
\left[\frac{67}{12}-\frac{\pi^2}{4}-\frac{5n_f}{18}-\frac{\beta_0}{2}\ln\frac{C_2}{C_1}\right].
\end{eqnarray}
%\begin{eqnarray}
%A=\frac{\alpha_s}{\pi}C_f
%+\frac{\alpha_s^2}{\pi^2}C_f
%\left[C_A\left(\frac{67}{36}-\frac{\pi^2}{12}\right)-\frac{5n_f}{18}-\frac{\beta_0}{2}\ln\frac{C_2}{C_1}\right],
%\end{eqnarray}
%\begin{eqnarray}
%&& A^{(1)}=C_f, \quad A^{(2)}=C_f\left[\frac{1}{2}K
%-\frac{\beta_0}{2}\ln\frac{C_2}{C_1}\right],
%\nonumber \\ &&
%K=C_A(\frac{67}{18}-\frac{\pi^2}{6})-\frac{5}{9}n_f, \quad {\beta_0}=11-\frac{2}{3}n_f,
%\end{eqnarray}
The color factor $C_f$ is
equal to  $C_F(=4/3)$ and $C_A(=3)$ for the light-quark and gluon jet, respectively,
$\beta_0$ is the QCD Beta function \cite{Amsler:2008zz}, and $n_f$ is the number
of active light-quark flavors.
The value of $\nu^2_{\rm in}$ diminishes the large logarithms in the
initial condition ${\bar J}_f^E(1,P_T,\nu^2_{\rm in},R,r)$,
which is then evaluated up to NLO including non-logarithmic-$r$ terms. The
value $\nu^2_{\rm fi}=1$ implies the presence of the large
logarithms in ${\bar J}_f^E(1,P_T,\nu^2_{\rm fi},R,r)$, which have
been summed into the Sudakov integral in Eq.~(\ref{ep1}).

We set  the ${\cal O}(1)$ constants $C_1=C_2=1$
and $C=\exp(5/2)$ ($C=\exp(17/6)$) for the quark (gluon)
jet in order to reproduce the large logarithms $\alpha_s\ln^2 r$ and
$\alpha_s\ln r$ in the NLO calculations.
The variation of these ${\cal
O}(1)$ constants reflects theoretical uncertainty  in our
formalism. The value of $r$ in the lower bound is taken to be larger than $0.1$, so
that it is safe to evaluate the Sudakov integral perturbatively. We
then derive the energy profile $\Psi(r)$ \cite{Acosta:2005ix} as the
energy fraction accumulated within the cone of size $r<R$ in terms
of the solution in Eq.~(\ref{ep1}), 
\begin{eqnarray}
\Psi(r)&=&\sum_f \int \frac{dP_T}{P_T}\frac{d\sigma_f}{dP_T} \bar
J_f^E(1, P_T, \nu_{\rm fi}^2, R,r)  \nonumber \\ &&\times
\left[\sum_f \int \frac{dP_T}{P_T}\frac{d\sigma_f}{dP_T} \bar
J_f^E(1, P_T, \nu_{\rm fi}^2, R,R)\right]^{-1} \, ,
 \label{3}
\end{eqnarray}
which respects the normalization $\Psi(r=R)=1$.
\begin{figure}[!htb]
\includegraphics[width=0.23\textwidth]{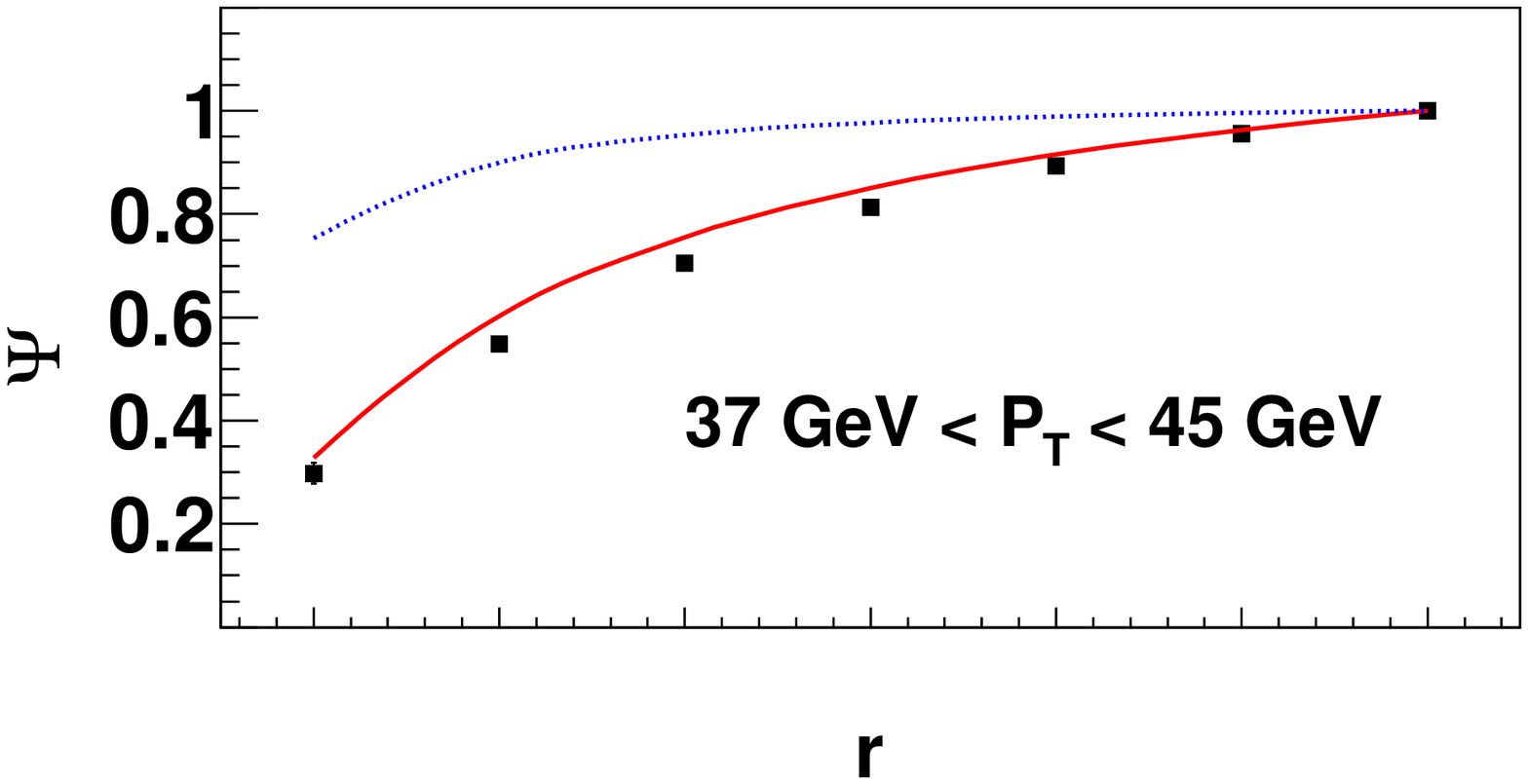}
\includegraphics[width=0.23\textwidth]{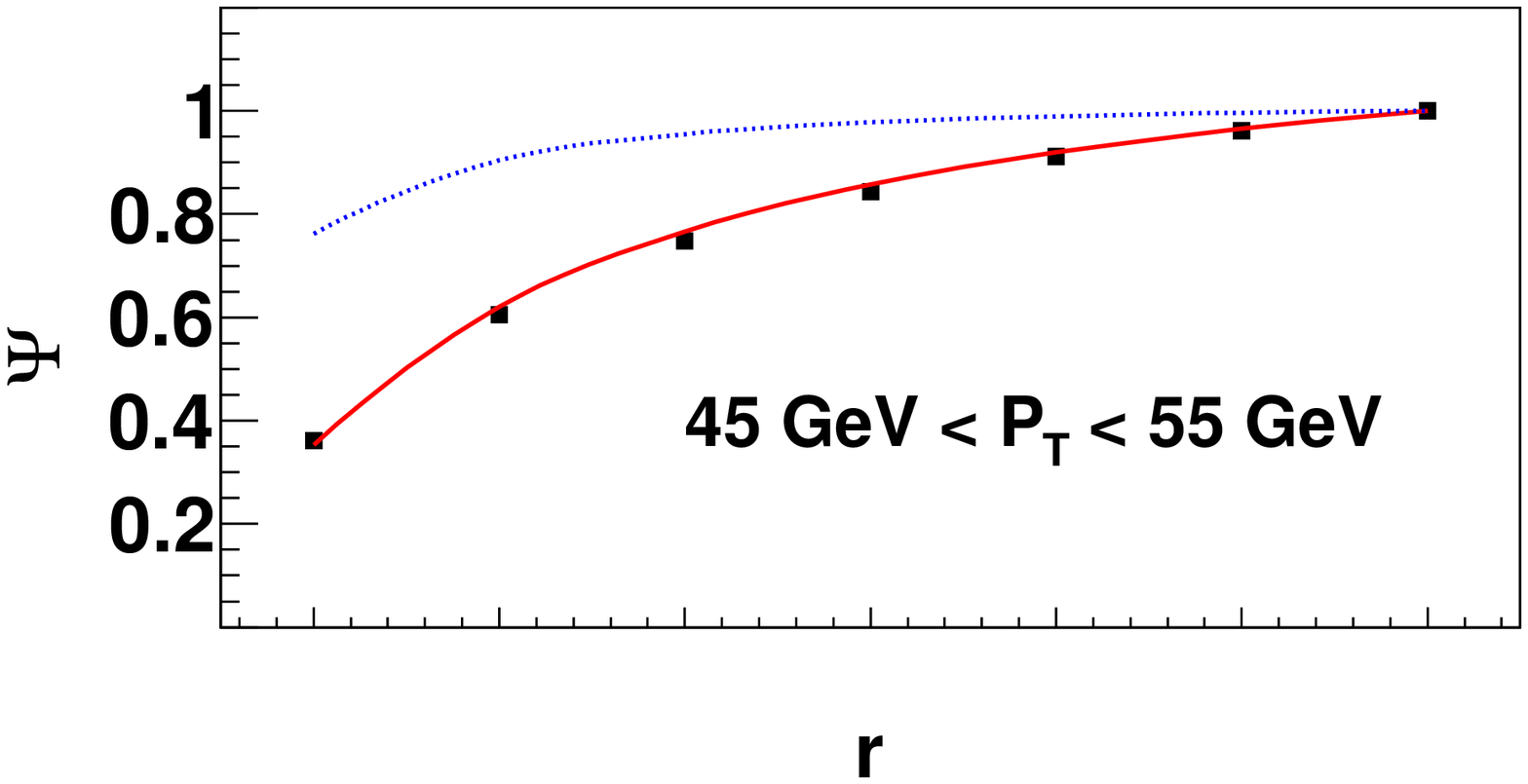}
\includegraphics[width=0.23\textwidth]{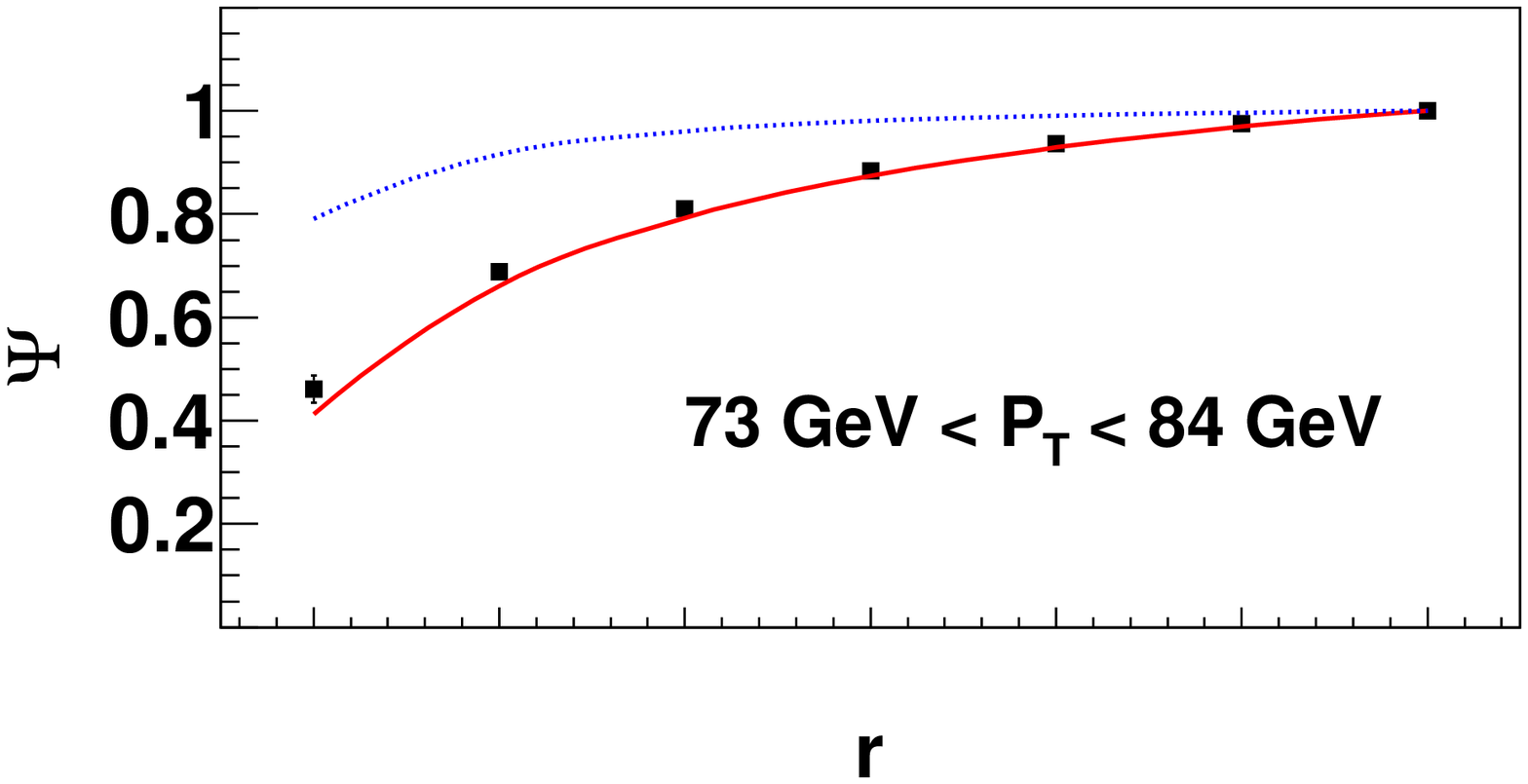}
\includegraphics[width=0.23\textwidth]{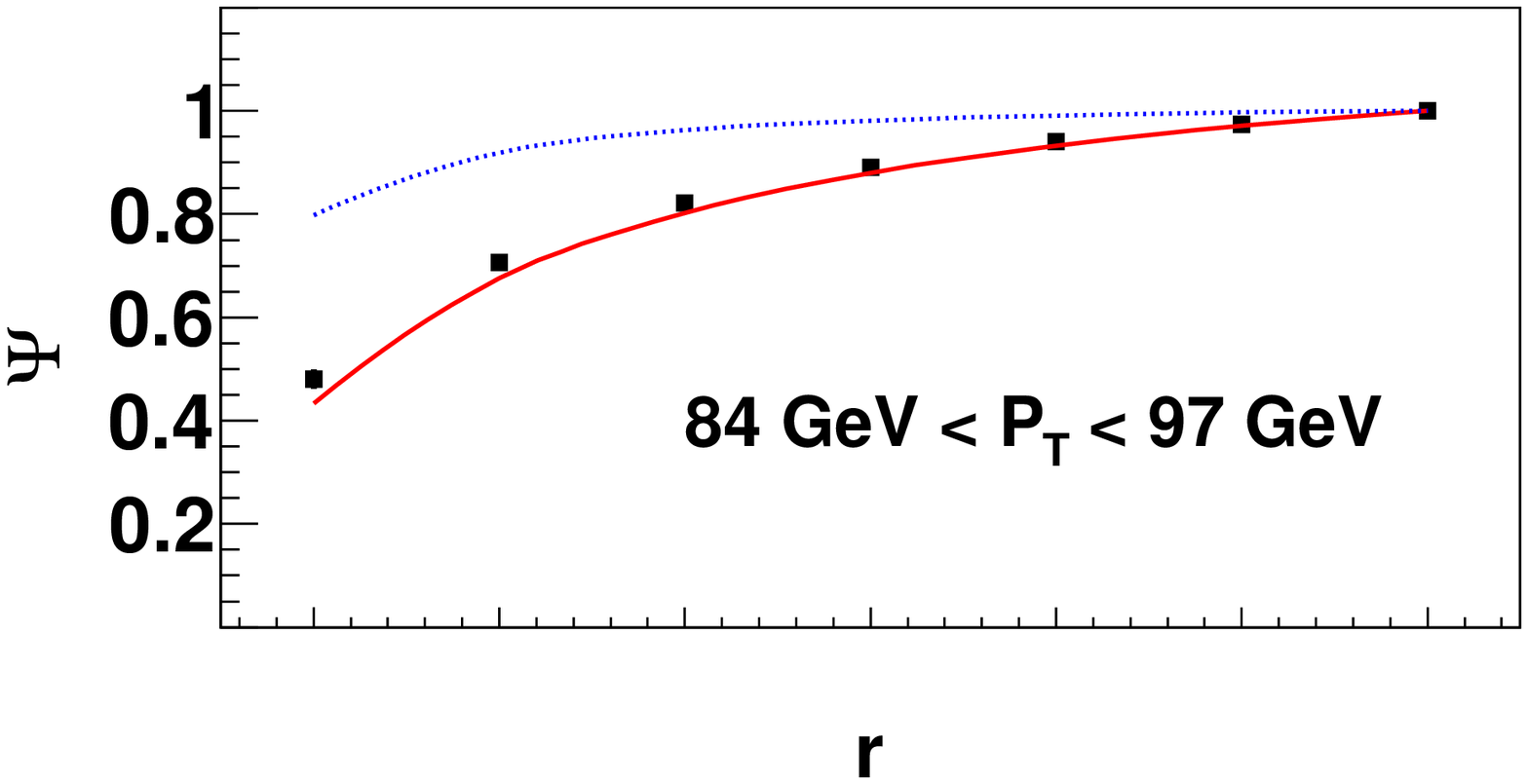}
\includegraphics[width=0.23\textwidth]{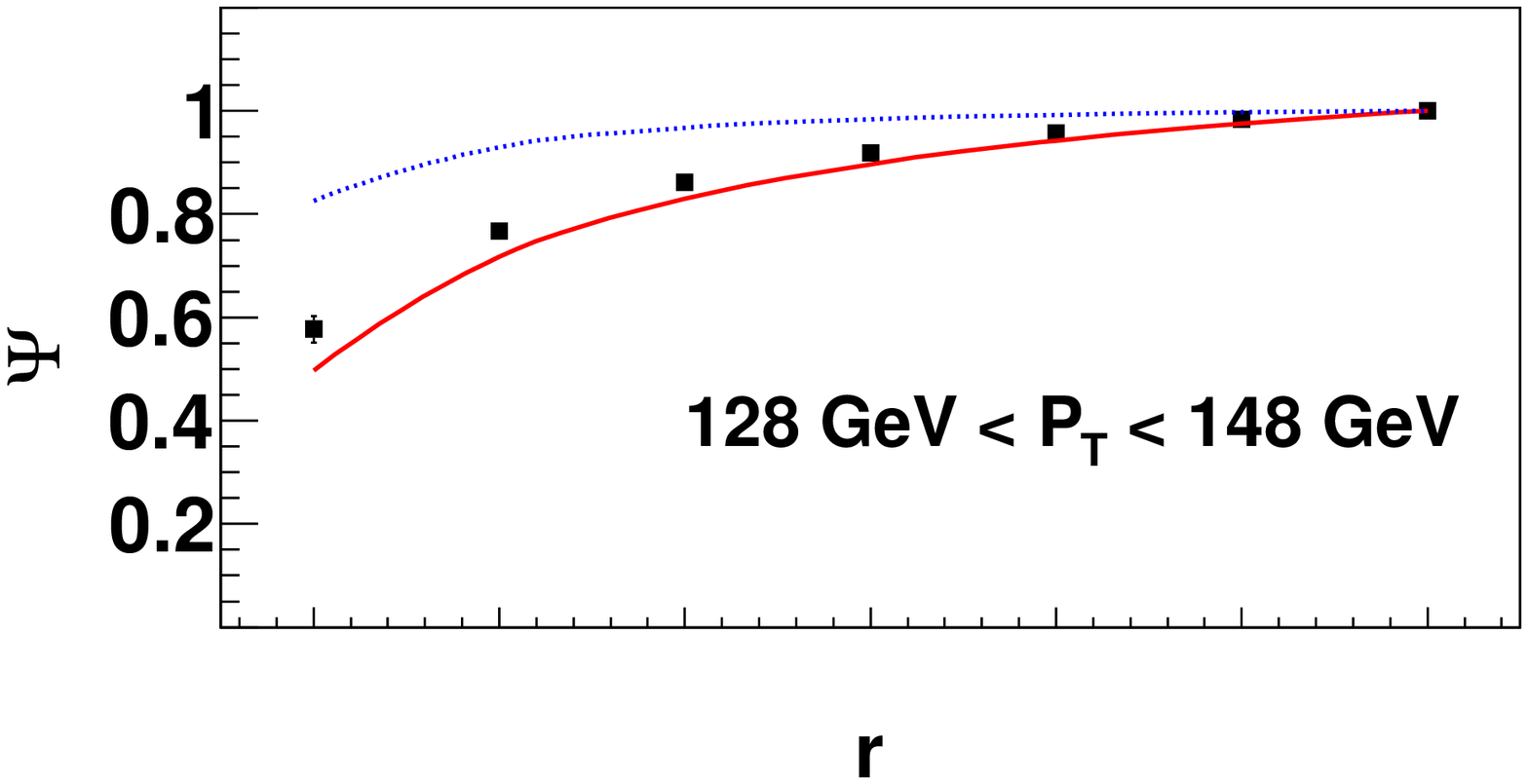}
\includegraphics[width=0.23\textwidth]{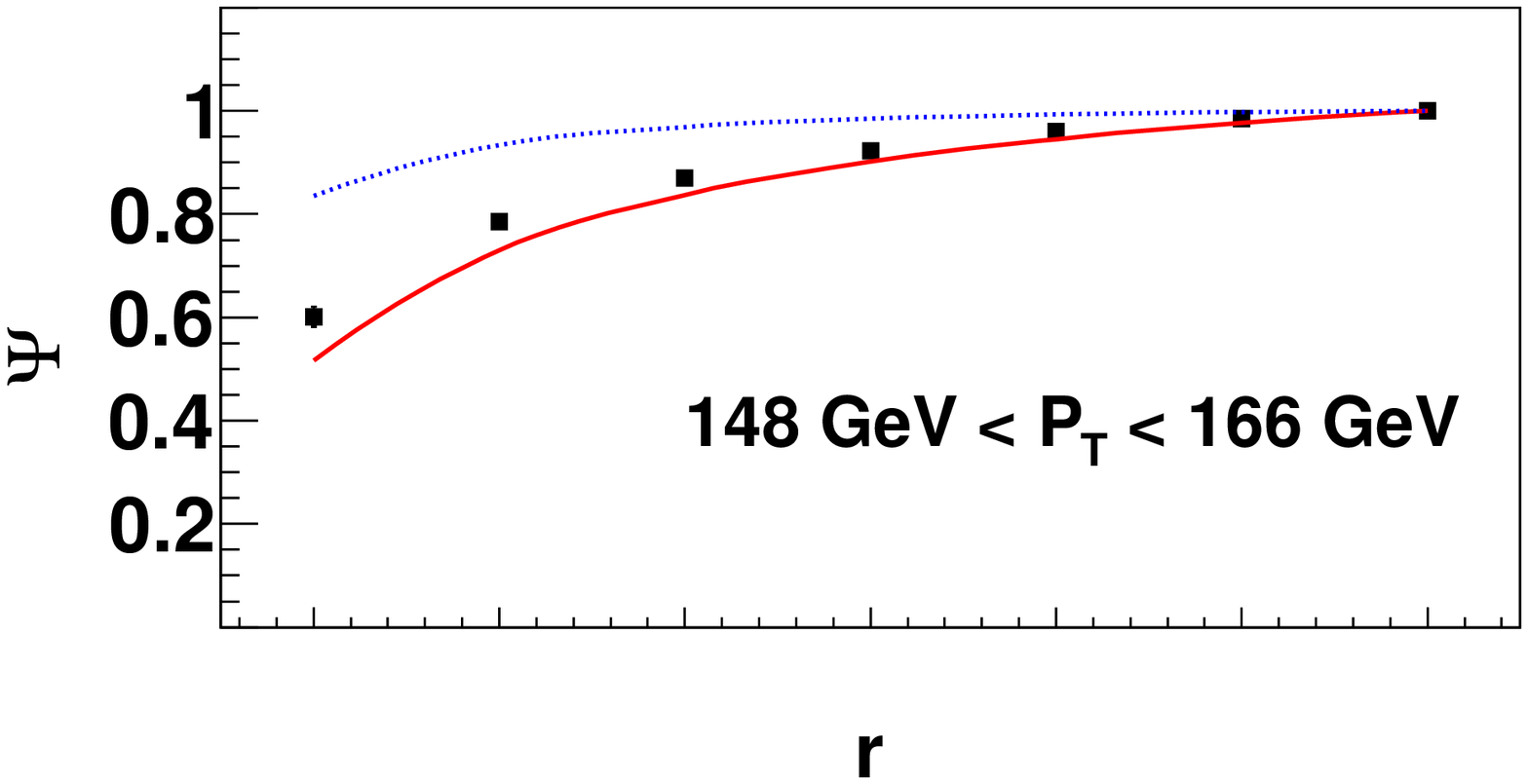}
\includegraphics[width=0.23\textwidth]{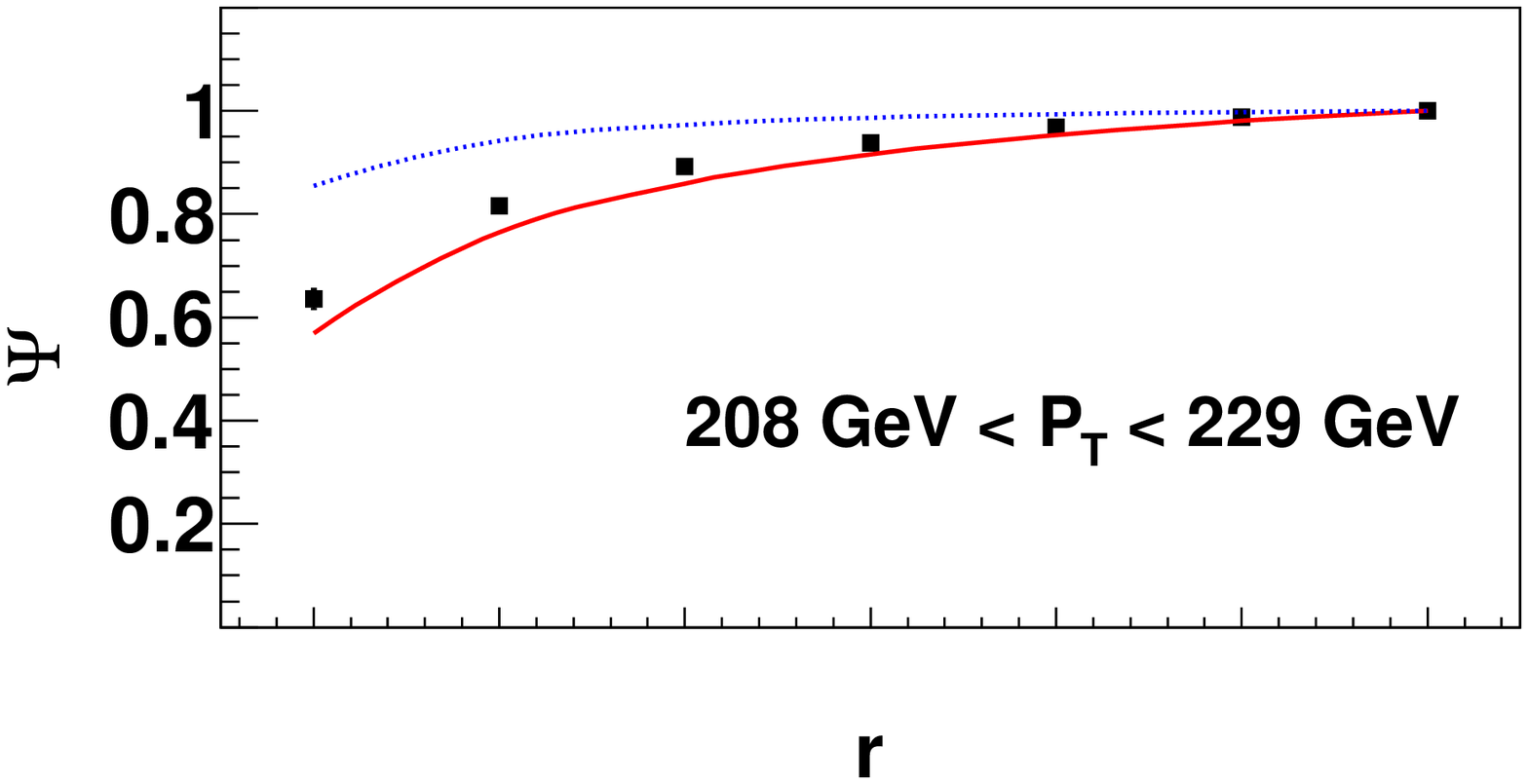}
\includegraphics[width=0.23\textwidth]{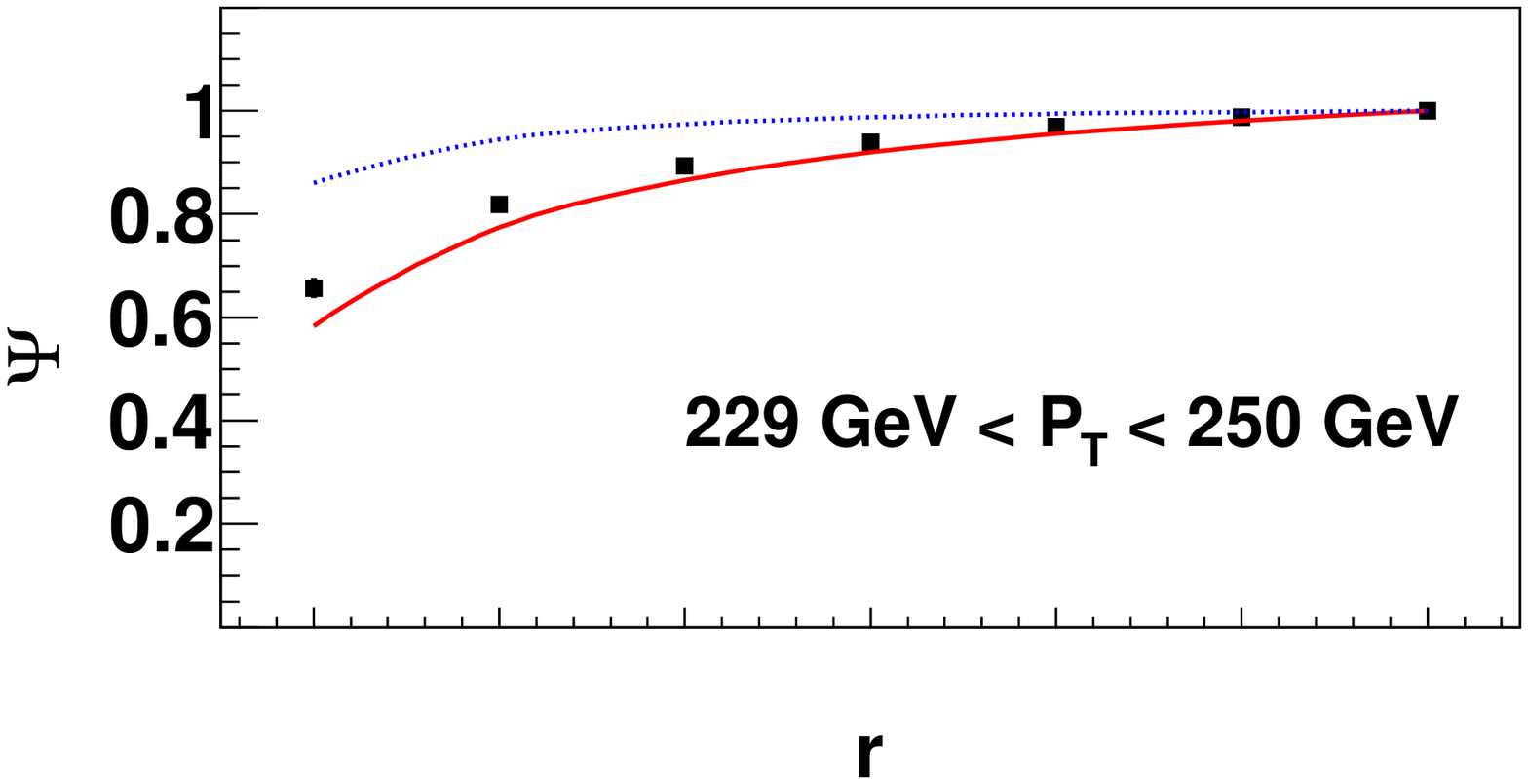}
\includegraphics[width=0.23\textwidth]{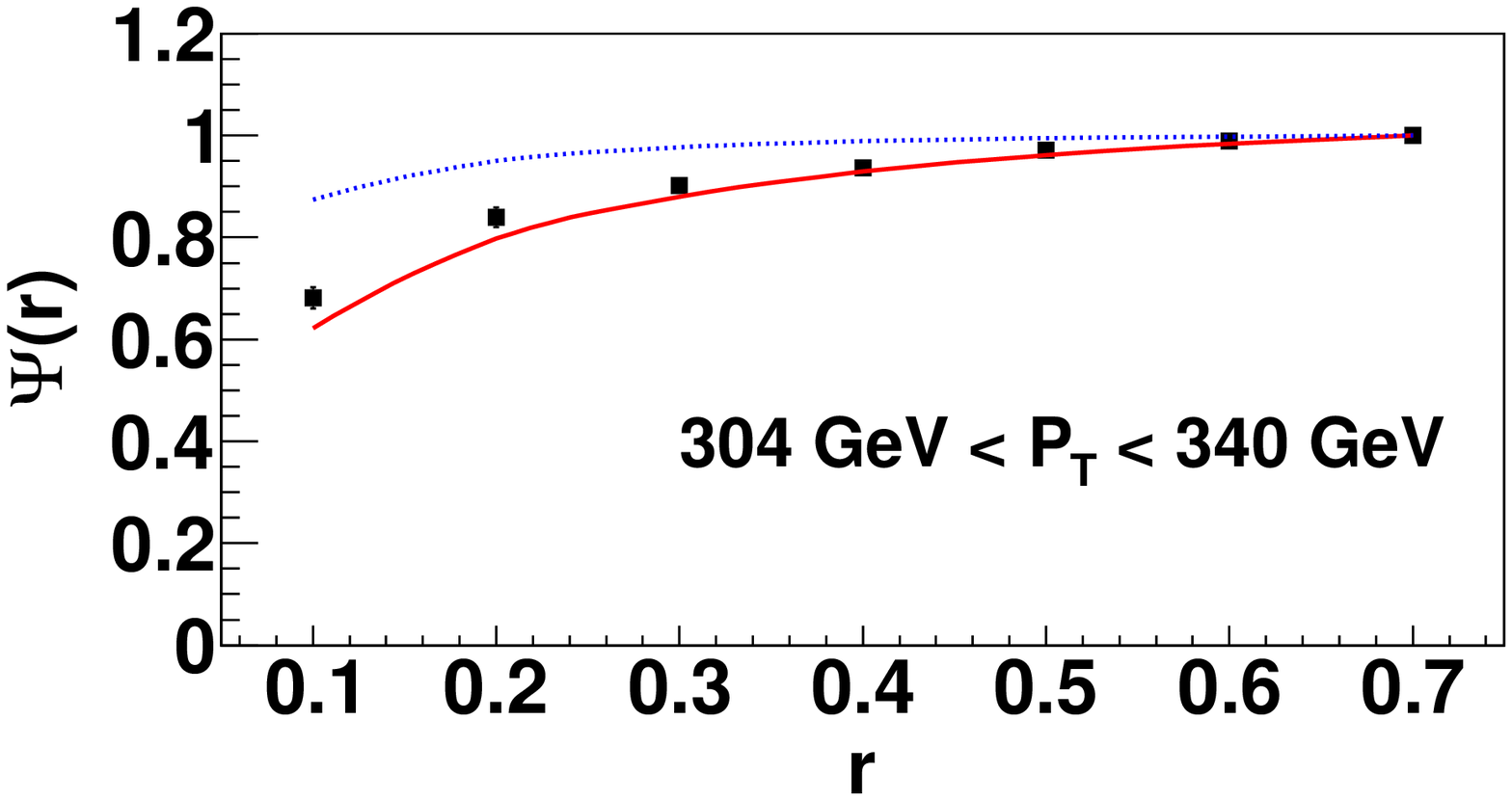}
\includegraphics[width=0.23\textwidth]{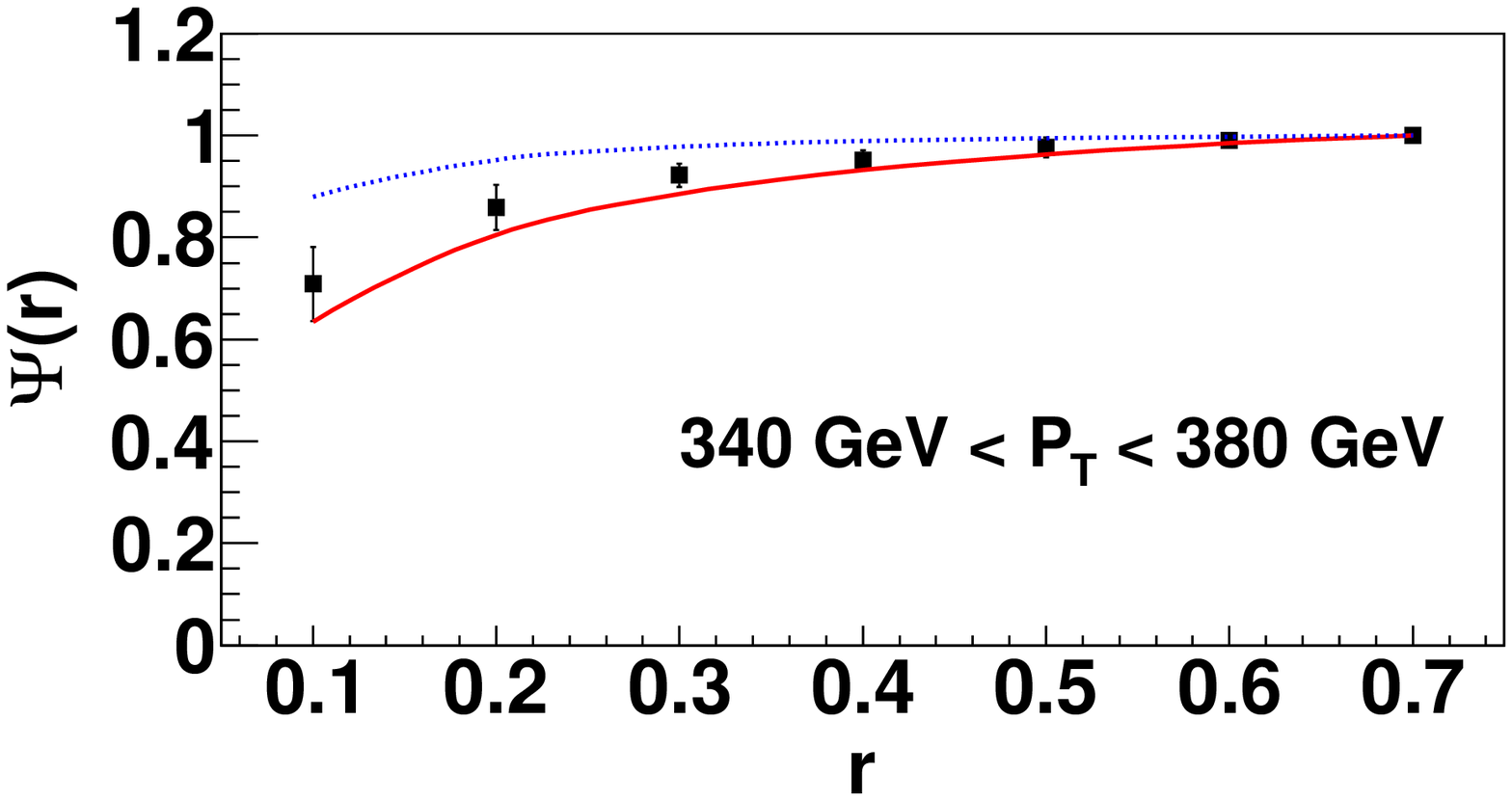}
\caption{Resummation (solid) and NLO (dashed) predictions for jet energy profiles compared
with CDF data \cite{Acosta:2005ix}.} \label{CDFJE}
\end{figure}
Equation (\ref{3}) contains the convolution of the LO differential
cross section $d\sigma_f/dP_T$ and the quark and gluon jet
energy functions.
Using the CTEQ6L parton distribution
functions (PDFs) \cite{Pumplin:2002vw}, we compare
the resummation and NLO predictions in Fig.~\ref{CDFJE} 
with the Tevatron CDF data \cite{Acosta:2005ix}.
The agreement between the resummation
predictions and the CDF data is obvious for all  $P_T$ values.
As $P_T$ increases, the accumulation of energy
inside the jets becomes faster.
\begin{figure}[!htb]
\includegraphics[width=0.23\textwidth]{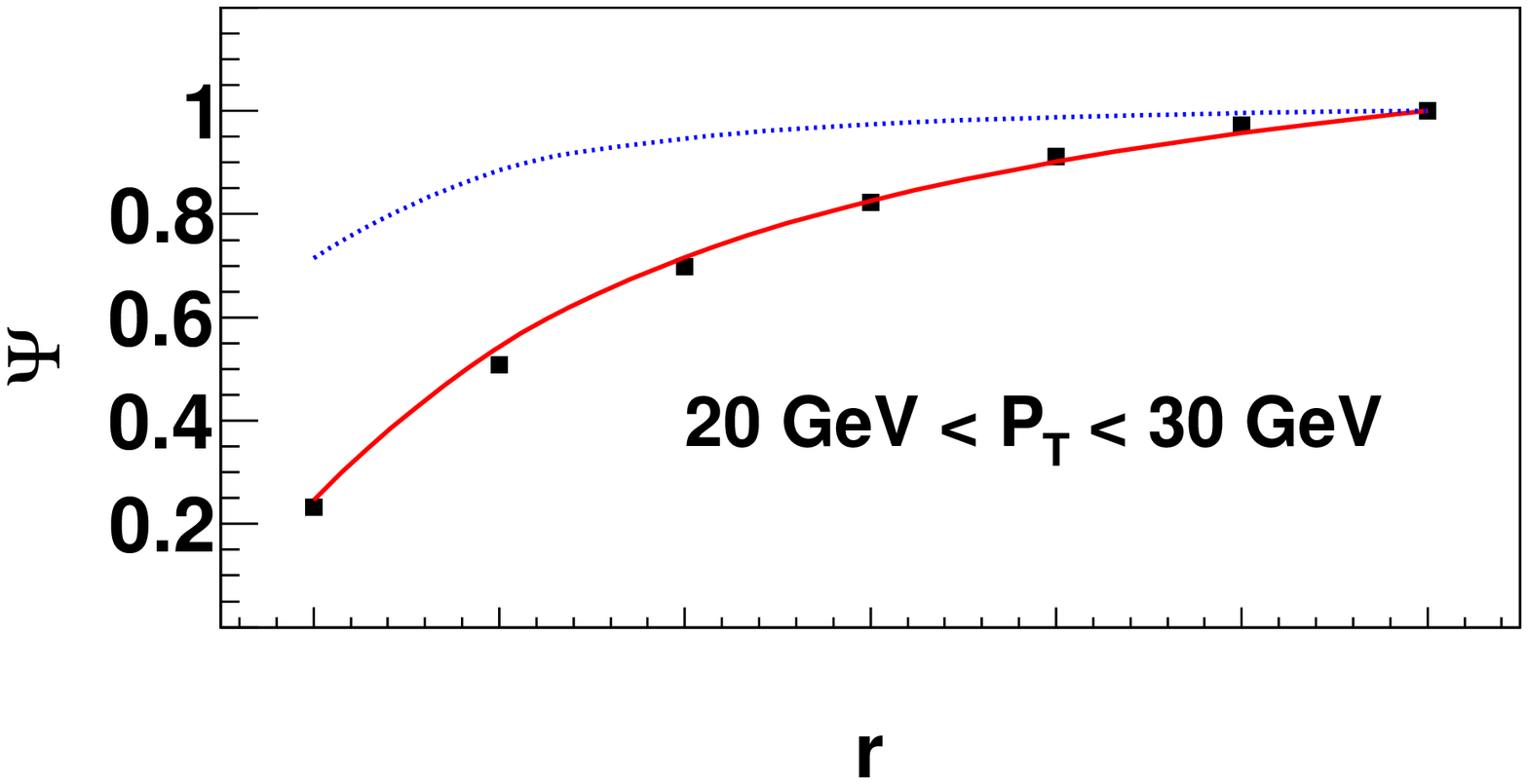}
\includegraphics[width=0.23\textwidth]{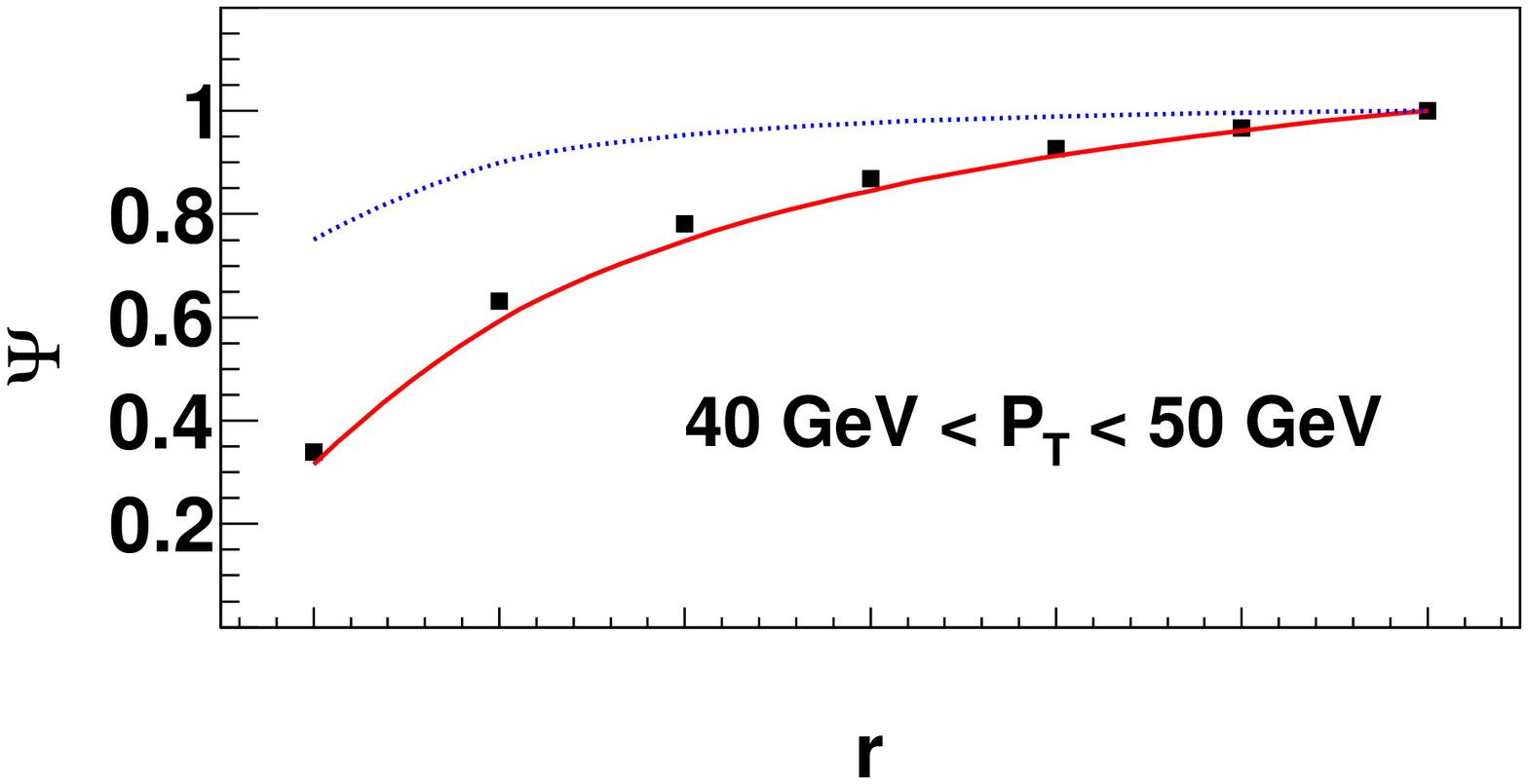}
\includegraphics[width=0.23\textwidth]{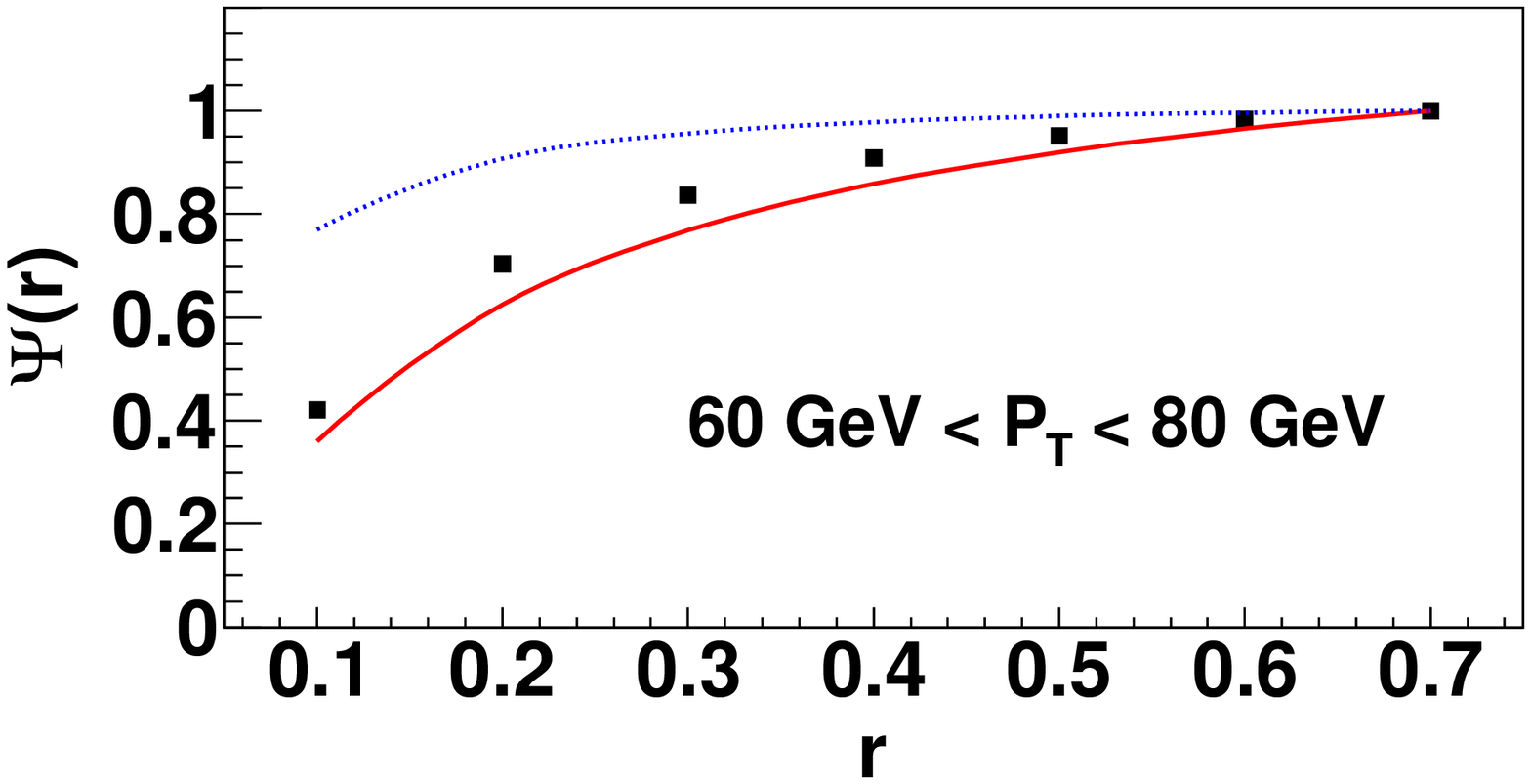}
\includegraphics[width=0.23\textwidth]{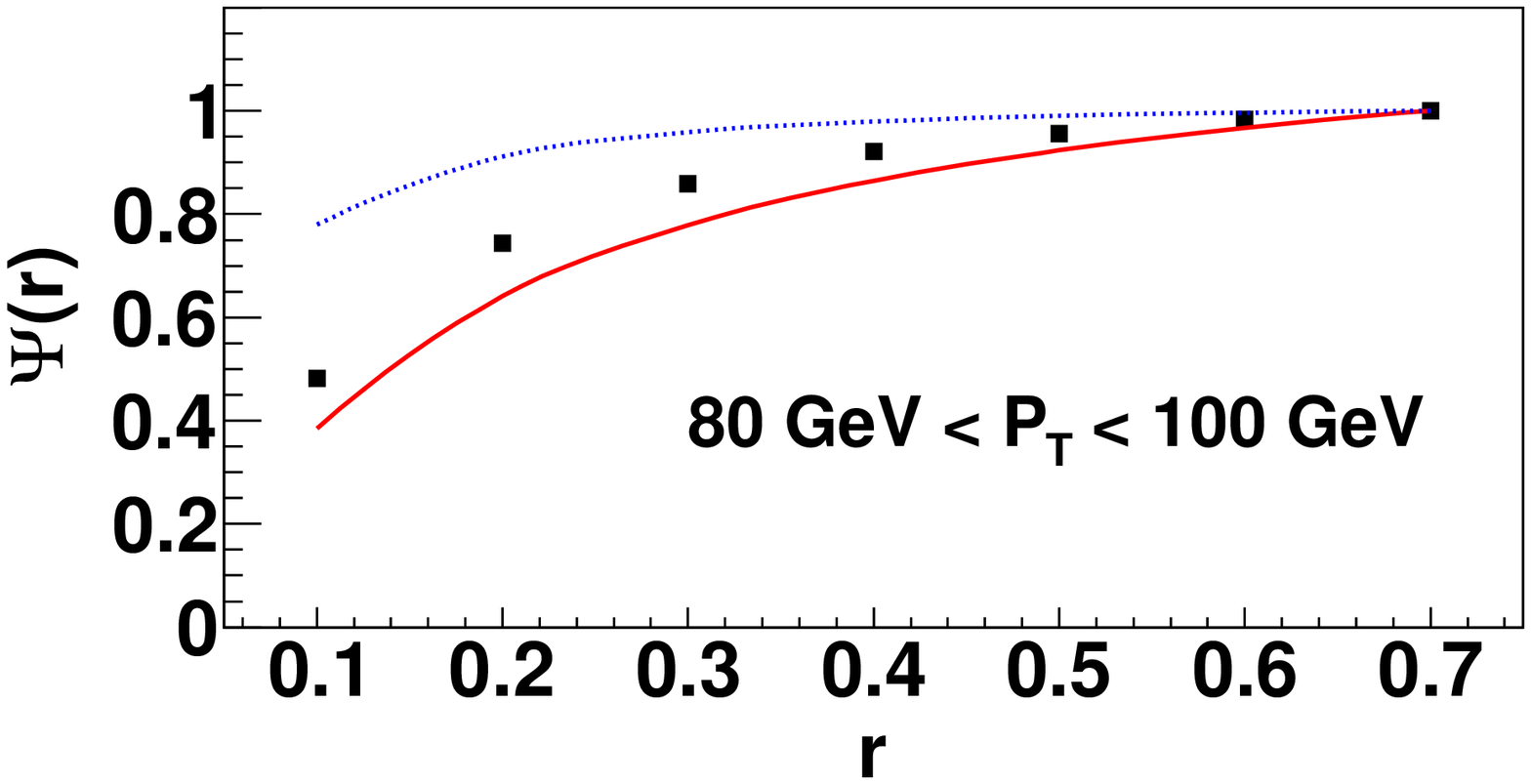}
\caption{Resummation (solid) and NLO (dashed) predictions for jet energy profiles compared
with CMS data \cite{CMSJE}.} \label{CMSJE}
\end{figure}
The NLO predictions derived from ${\bar J}_f^{E(1)}(1,P_T,\nu^2_{\rm fi},R,r)$
are also displayed, which overshoot the data.
Figure~\ref{CMSJE} shows the agreement of the resummation predictions for the
jet energy profiles with the LHC CMS data at 7 TeV \cite{CMSJE} and
the overshooting of the NLO predictions.
The above consistency indicates that our resummation formalism has
captured the dominant dynamics in a jet energy profile and can give
a direct and reliable prediction for this observable.

As stated before, the invariant mass distribution of an energetic
jet can be utilized as an experimental signature of new physics. We shall
demonstrate below that our formalism is also applicable to this jet
substructure by predicting the mass distributions of the light-quark
and gluon jets.
Following the previous analysis on jet energy profiles,
we vary the Wilson line into an arbitrary
direction $n^\mu$ with $n^2\not=0$, when implementing the
resummation technique \cite{Li:1995eh} to derive the evolution
equation for the jet function $J_f(M_J^2,P_T,\nu^2,R)$. In this
case, the overlap of the collinear and soft enhancements generate
the double logarithms of the ratio $(P_J\cdot n)^2/(M_J^2n^2)\equiv
(R^2P_T^2/(4M_J^2))\nu^2$, so that the variation of $n$, i.e.,
$\nu^2$, can turn into the variation of $M_J$. The dependence on the
jet cone size $R$ is introduced through the Mellin transformation
with respect to the variable $x \equiv  M_J^2/(R^2P_T^2)$.
The solution describing the evolution of the jet function from the
initial value $\nu^2_{\rm in}=C_1/(C_2\bar N)$, where $\bar N = N
\exp(\gamma_E)$ with $\gamma_E$ being the Euler's constant, to the
final value $\nu^2_{\rm fi}=1$ is given by
\begin{eqnarray}
{\bar J}_f(N,P_T,\nu^2_{\rm fi},R)&=&{\bar J}_f(N,P_T,\nu^2_{\rm
in},R) \exp[S(N)], \label{s1}
\end{eqnarray}
in which the Sudakov exponent is written as
\begin{eqnarray}
S(N)&=&-\int_{C_1/\bar N}^{C_2}\frac{dy}{y}\left\{
A(\alpha_s(y^2 R^2 P_T^2))\ln\left(\frac{C_2}{y}\right)
\right. \nonumber \\ && \left.
-\frac{C_f}{\pi}\alpha_s\left(y^2R^2P_T^2\right)
\left[\frac{1}{2}+\ln\frac{C_2}{C_1}\right]\right\}.
\label{sud}
\end{eqnarray}
The value $\nu^2_{\rm fi}=1$ implies the presence of the large
logarithms (in powers of $\ln N$) in ${\bar J}_f(N,P_T,\nu^2_{\rm fi},R)$, which
have been summed into the Sudakov exponent. The initial condition
${\bar J}_f(N,P_T,\nu^2_{\rm in},R)$ can then be evaluated
perturbatively. At NLO, they have the expressions
\begin{eqnarray}
&&\bar J_q^{(1)}=\frac{1}{R^2 P_T^2}\left\{
1+\frac{C_F}{\pi}\alpha_s\left(C_2^2 R^2 P_T^2/C_1^2\right) \left[
\frac{1}{2}\ln\frac{C_1}{C_2} \right.\right. \nonumber \\  &&
\left.\left. -\frac{1}{2}\ln^2\frac{C_1}{C_2}
+\frac{1}{2}\gamma_E-\frac{\pi^2}{4}-\frac{9}{8} \right]\right\},\label{qj}\\
&&\bar J_g^{(1)}= \frac{1}{R^2 P_T^2}\left\{
1+\frac{C_A}{\pi}\alpha_s\left(C_2^2 R^2 P_T^2/C_1^2\right) \left[
\frac{1}{2}\ln\frac{C_1}{C_2} \right.\right. \nonumber \\ &&
\left.\left. -\frac{1}{2}\ln^2\frac{C_1}{C_2}
-\frac{5}{12}\gamma_E-\frac{\pi^2}{4}+\frac{1}{2}(\ln
2-3)+\frac{1}{36} \right]\right\}
\label{gj}
\end{eqnarray}
for the light-quark and gluon jets, respectively. The remaining
$N$-dependent terms are suppressed by $1/N$ and their effect is
small.

For a fixed $P_T$, the scale $R P_T/N$ involved in Eq.~(\ref{sud})
becomes so low at extremely large $N$ that the perturbative analysis
fails and  nonperturbative contributions, arising from hadronization and
underlying events, need to be included in order to predict jet mass
distribution in the small $M_J$ region. For convenience, we
introduce
\begin{equation}
S^{NP}(N)= \frac{N^2 Q_0^2 }{R^2P_T^2}(C_f \alpha_0\ln N
+\alpha_1)+C_f\alpha_2\frac{N Q_0}{R P_T},\label{np}
\end{equation}
into the Sudakov exponent with $Q_0$ being set to $1$ GeV.
It consists of a logarithmic term and a Gaussian
smearing term (as suggested by $S(N)$ in Eq.~(\ref{sud})),
 as well as a linear term in $N$ (for describing a
final-state jet \cite{Dasgupta:1998eqa}).
The parameters $\alpha_{0,1,2}$ can be determined by a fit to
experimental data for certain jet momentum and jet cone. In this
work, we perform fits to full event generators PYTHIA
\cite{Sjostrand:2007gs} and SpartyJet \cite{Ellis:2007ib} for the
quark (gluon) jet produced at the Tevatron run 2 energy $1.96$ TeV
with $P_T=600$ GeV and $R=0.7$, which yields $\alpha_0=-0.35$,
$\alpha_1=0.50$ ($-4.59$), and $\alpha_2=-1.66$.
With this nonperturbative contribution, we are ready to predict
jet mass distribution for arbitrary values of $P_T$ and $R$ using
the improved resummation solution:
\begin{eqnarray}
{\bar J}_f^{\rm RES}(N,P_T,\nu^2_{\rm fi},R)&=&{\bar
J}_f(N,P_T,\nu^2_{\rm in},R) \nonumber \\  && \times
\exp\left[S(N)+S^{NP}(N)\right]  .
\end{eqnarray}

The jet function in $M_J$ space is derived via the inverse Mellin
transformation
\begin{eqnarray}
J_f^{\rm RES} (M_J^2) = \frac{1}{2\pi i} \int_{\rm C} dN
(1-x)^{-N} {\bar J}_f^{\rm RES} (N),
\end{eqnarray}
where $x=M_J^2/(R^2P_T^2)$ and the contour $\rm C$ runs along the negative real
axis in the complex-$N$ plane and circles the origin  counterclockwise with a finite radius.
To avoid the Landau pole, we flatten the running behavior of
$\alpha_s$ \cite{Vogt:2000ci,Amsler:2008zz} at certain scale
$\mu_c$:
\begin{eqnarray}
\alpha_s(\mu)=\left\{
\begin{array}{ll}
\alpha_s(\mu_c\exp[{\rm i}{\rm Arg}(\mu)]), & |\mu|<\mu_c
\\
\alpha_s(\mu), & |\mu|>\mu_c
\end{array}
\right.. \label{alphasCut}
\end{eqnarray}
In the small $M_J$ region, all moments in $N$ are equally important,
and those containing powers of $\ln N$ dominate. Therefore, our resummation
formalism, in which these large logarithms are summed up, can give
reliable predictions. In the large $M_J$ region, the nonlogarithmic terms are
not negligible, so we have to improve the resummation formula by
including the nonlogarithmic terms (say, up to NLO), namely, by matching
fixed-order results to the resummation formula as done for the Drell-Yan
processes \cite{Balazs:1997xd}.

\begin{figure}[!htb]
\includegraphics[width=0.35\textwidth]{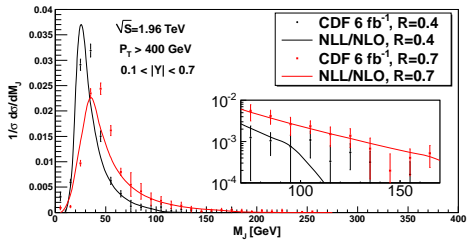}
\caption{Comparison between resummation predictions and Tevatron
data.} \label{CONVP1}
\end{figure}

\begin{figure}[!htb]
\includegraphics[width=0.35\textwidth]{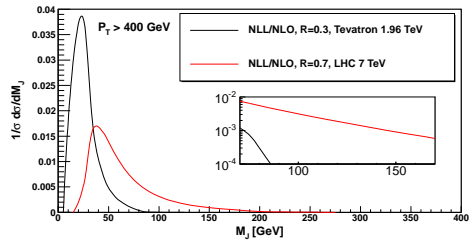}
\caption{Resummation predictions for jet mass distributions at
Tevatron and LHC.} \label{CONVP2}
\end{figure}

The jet mass distribution is calculated by convoluting the LO
differential cross section of inclusive (quark or gluon) jet
production with the corresponding quark or gluon jet function,
\begin{eqnarray}
\frac{d\sigma} {dM_J^2}
&=&\sum_f \int dP_T \frac{d\sigma_f}{dP_T} J_f(M_J^2, P_T,\nu_{\rm fi}^2,R).
\end{eqnarray}
We are now ready to compare our predictions for the jet mass
distributions with the Tevatron and LHC data, by choosing
$C_1=\exp(\gamma_E)$, $C_2=\exp(-\gamma_E)$ and $\mu_c=0.3$ GeV in
the numerical analysis with  the CTEQ6L PDFs \cite{Pumplin:2002vw}.
Results from different choices of $C_1$,
$C_2$ and $\mu_c$ reveal theoretical uncertainty, which will be
investigated in a forthcoming paper.
The comparison to the Tevatron data
\cite{CDFJM}, with the kinematic cuts $P_T>400$ GeV and $0.1<|Y|<0.7$,
is presented  in Fig.~\ref{CONVP1}, where the label NLL/NLO denotes
the prediction of NLL resummation with the NLO initial conditions given in
Eqs.~(\ref{qj}) and (\ref{gj}). It shows  that our predictions
agree well with the CDF data for $R=0.4$ and $0.7$ in the intermediate
jet mass region. In the small invariant mass region, the predictions can
describe the overall shape of the jet distributions within the
uncertainty induced by PDFs \cite{CDFJM}, though the peak positions
are slightly lower than the CDF data.
However, for large $M_J$, e.g. $M_J > 100 (200)$ GeV for $R = 0.4 (0.7)$, 
the resummation prediction drops off quickly and deviates from data. 
This can be further improved by matching to exact calculations of NLO and beyond.
The resummation predictions for the jet mass distributions at
Tevatron with $R=0.3$ and at LHC with $R=0.7$ are shown in
Fig.~\ref{CONVP2}, which can be tested by future measurements.

In conclusion, we have developed a theoretical framework based on
the pQCD theory for analyzing the substructures of the light-quark
and gluon jets. This is the first time in the literature that pQCD
is shown to describe well the jet energy profiles and mass
distributions, which are the most commonly discussed physical
observables to describe the substructure of a jet signal at
hadron colliders. We have demonstrated
that the resummation predictions for the jet energy profiles, with
the jet invariant mass being integrated out, are insensitive to
nonperturbative inputs, and in excellent agreement with
both the Tevatron CDF and LHC CMS data,
for arbitrary values of  jet momentum. In view of the fact that the jet
energy profile is a useful feature for jet identification at the
LHC, the energy profiles of boosted top-quark jets, and jets from
boosted Higgs boson and weak gauge boson ($W$, $Z$ or $Z'$, {\it etc.})
hadronic decays will be studied in a forthcoming paper. We shall
also calculate other jet substructures, such as the angularity
\cite{Ellis:2010rwa}. These studies are
crucial for the LHC physics program in terms of testing the QCD theory
and identifying new physics signals. We have also applied the
resummation formalism to predict the light-quark and gluon jet mass
distributions. To describe jet mass distributions in the low mass
region, we need to introduce some nonperturbative contributions in
the resummation formalism. The relevant parameters can be fixed at
one jet energy scale, and then employed to make predictions for
other energy scales, which are found to agree with the Tevatron
data. It is expected that our formalism will successfully apply to upcoming LHC
data of jet mass distributions.

\acknowledgments{

This work was supported by National Science Council of R.O.C. under
Grant No. NSC 98-2112-M-001-015-MY3, and by the U.S. National
Science Foundation under Grant No. PHY-0855561. CPY and ZL acknowledge the
hospitality of Academia Sinica and National Center for Theoretical
Sciences in Taiwan, where part of this work was done. We thank Pekka
Sinervo and Raz Alon for providing CDF jet
mass distribution data.}

\bibliography{shortlightjet1}

\begin{thebibliography}{24}
\expandafter\ifx\csname natexlab\endcsname\relax\def\natexlab#1{#1}\fi
\expandafter\ifx\csname bibnamefont\endcsname\relax
  \def\bibnamefont#1{#1}\fi
\expandafter\ifx\csname bibfnamefont\endcsname\relax
  \def\bibfnamefont#1{#1}\fi
\expandafter\ifx\csname citenamefont\endcsname\relax
  \def\citenamefont#1{#1}\fi
\expandafter\ifx\csname url\endcsname\relax
  \def\url#1{\texttt{#1}}\fi
\expandafter\ifx\csname urlprefix\endcsname\relax\def\urlprefix{URL }\fi
\providecommand{\bibinfo}[2]{#2}
\providecommand{\eprint}[2][]{\url{#2}}

\bibitem[{\citenamefont{Ellis et~al.}(2010)\citenamefont{Ellis, Vermilion,
  Walsh, Hornig, and Lee}}]{Ellis:2010rwa}
\bibinfo{author}{\bibfnamefont{S.~D.} \bibnamefont{Ellis}},
  \bibinfo{author}{\bibfnamefont{C.~K.} \bibnamefont{Vermilion}},
  \bibinfo{author}{\bibfnamefont{J.~R.} \bibnamefont{Walsh}},
  \bibinfo{author}{\bibfnamefont{A.}~\bibnamefont{Hornig}}, \bibnamefont{and}
  \bibinfo{author}{\bibfnamefont{C.}~\bibnamefont{Lee}},
  \bibinfo{journal}{JHEP} \textbf{\bibinfo{volume}{1011}}, \bibinfo{pages}{101}
  (\bibinfo{year}{2010}).

\bibitem[{\citenamefont{Kelley et~al.}(2011)\citenamefont{Kelley, Schwartz, and
  Zhu}}]{Kelley:2011tj}
\bibinfo{author}{\bibfnamefont{R.}~\bibnamefont{Kelley}},
  \bibinfo{author}{\bibfnamefont{M.~D.} \bibnamefont{Schwartz}},
  \bibnamefont{and} \bibinfo{author}{\bibfnamefont{H.~X.} \bibnamefont{Zhu}}
  (\bibinfo{year}{2011}), \eprint{1102.0561}.

\bibitem[{\citenamefont{Collins et~al.}(1985)\citenamefont{Collins, Soper, and
  Sterman}}]{Collins:1984kg}
\bibinfo{author}{\bibfnamefont{J.~C.} \bibnamefont{Collins}},
  \bibinfo{author}{\bibfnamefont{D.~E.} \bibnamefont{Soper}}, \bibnamefont{and}
  \bibinfo{author}{\bibfnamefont{G.~F.} \bibnamefont{Sterman}},
  \bibinfo{journal}{Nucl.Phys.} \textbf{\bibinfo{volume}{B250}},
  \bibinfo{pages}{199} (\bibinfo{year}{1985}).

\bibitem[{\citenamefont{Acosta et~al.}(2005)}]{Acosta:2005ix}
\bibinfo{author}{\bibfnamefont{D.~E.} \bibnamefont{Acosta}}
  \bibnamefont{et~al.} (\bibinfo{collaboration}{CDF Collaboration}),
  \bibinfo{journal}{Phys. Rev.} \textbf{\bibinfo{volume}{D71}},
  \bibinfo{pages}{112002} (\bibinfo{year}{2005}).

\bibitem[{CMS(2010)}]{CMSJE}
\bibinfo{type}{Report} \bibinfo{number}{CMS--PAS--QCD--10--014}
  (\bibinfo{year}{2010}).

\bibitem[{CDF(2011)}]{CDFJM}
\bibinfo{type}{Report} \bibinfo{number}{CDF--PUB--JET--PUBLIC--10119}
  (\bibinfo{year}{2011}).

\bibitem[{\citenamefont{Agashe et~al.}(2008)}]{Agashe:2006hk}
\bibinfo{author}{\bibfnamefont{K.}~\bibnamefont{Agashe}} \bibnamefont{et~al.},
  \bibinfo{journal}{Phys. Rev.} \textbf{\bibinfo{volume}{D77}},
  \bibinfo{pages}{015003} (\bibinfo{year}{2008}).

\bibitem[{\citenamefont{Fitzpatrick et~al.}(2007)}]{Fitzpatrick:2007qr}
\bibinfo{author}{\bibfnamefont{A.~L.} \bibnamefont{Fitzpatrick}}
  \bibnamefont{et~al.}, \bibinfo{journal}{JHEP} \textbf{\bibinfo{volume}{09}},
  \bibinfo{pages}{013} (\bibinfo{year}{2007}).

\bibitem[{\citenamefont{Baur and Orr}(2007)}]{Baur:2007ck}
\bibinfo{author}{\bibfnamefont{U.}~\bibnamefont{Baur}} \bibnamefont{and}
  \bibinfo{author}{\bibfnamefont{L.~H.} \bibnamefont{Orr}},
  \bibinfo{journal}{Phys. Rev.} \textbf{\bibinfo{volume}{D76}},
  \bibinfo{pages}{094012} (\bibinfo{year}{2007}).

\bibitem[{\citenamefont{Brooijmans et~al.}(2008)}]{Brooijmans:2008se}
\bibinfo{author}{\bibfnamefont{G.~H.} \bibnamefont{Brooijmans}}
  \bibnamefont{et~al.}, pp. \bibinfo{pages}{363--489} (\bibinfo{year}{2008}),
  \eprint{0802.3715}.

\bibitem[{\citenamefont{Skiba and Tucker-Smith}(2007)}]{Skiba:2007fw}
\bibinfo{author}{\bibfnamefont{W.}~\bibnamefont{Skiba}} \bibnamefont{and}
  \bibinfo{author}{\bibfnamefont{D.}~\bibnamefont{Tucker-Smith}},
  \bibinfo{journal}{Phys. Rev.} \textbf{\bibinfo{volume}{D75}},
  \bibinfo{pages}{115010} (\bibinfo{year}{2007}).

\bibitem[{\citenamefont{Holdom}(2007)}]{Holdom:2007ap}
\bibinfo{author}{\bibfnamefont{B.}~\bibnamefont{Holdom}},
  \bibinfo{journal}{JHEP} \textbf{\bibinfo{volume}{08}}, \bibinfo{pages}{069}
  (\bibinfo{year}{2007}).

\bibitem[{\citenamefont{Butterworth et~al.}(2008)}]{Butterworth:2008iy}
\bibinfo{author}{\bibfnamefont{J.~M.} \bibnamefont{Butterworth}}
  \bibnamefont{et~al.}, \bibinfo{journal}{Phys. Rev. Lett.}
  \textbf{\bibinfo{volume}{100}}, \bibinfo{pages}{242001}
  (\bibinfo{year}{2008}).

\bibitem[{\citenamefont{Gabrielli et~al.}(2007)}]{Gabrielli:2007wf}
\bibinfo{author}{\bibfnamefont{E.}~\bibnamefont{Gabrielli}}
  \bibnamefont{et~al.}, \bibinfo{journal}{Nucl. Phys.}
  \textbf{\bibinfo{volume}{B781}}, \bibinfo{pages}{64} (\bibinfo{year}{2007}).

\bibitem[{\citenamefont{Almeida et~al.}(2009)}]{Almeida:2008tp}
\bibinfo{author}{\bibfnamefont{L.~G.} \bibnamefont{Almeida}}
  \bibnamefont{et~al.}, \bibinfo{journal}{Phys. Rev.}
  \textbf{\bibinfo{volume}{D79}}, \bibinfo{pages}{074012}
  (\bibinfo{year}{2009}).

\bibitem[{\citenamefont{Li and Yu}(1996)}]{Li:1995eh}
\bibinfo{author}{\bibfnamefont{H.-n.} \bibnamefont{Li}} \bibnamefont{and}
  \bibinfo{author}{\bibfnamefont{H.-L.} \bibnamefont{Yu}},
  \bibinfo{journal}{Phys. Rev.} \textbf{\bibinfo{volume}{D53}},
  \bibinfo{pages}{4970} (\bibinfo{year}{1996}).

\bibitem[{\citenamefont{Li}(1997)}]{Li:1996gi}
\bibinfo{author}{\bibfnamefont{H.-n.} \bibnamefont{Li}},
  \bibinfo{journal}{Phys. Rev.} \textbf{\bibinfo{volume}{D55}},
  \bibinfo{pages}{105} (\bibinfo{year}{1997}).

\bibitem[{\citenamefont{Amsler et~al.}(2008)}]{Amsler:2008zz}
\bibinfo{author}{\bibfnamefont{C.}~\bibnamefont{Amsler}} \bibnamefont{et~al.}
  (\bibinfo{collaboration}{Particle Data Group}), \bibinfo{journal}{Phys.
  Lett.} \textbf{\bibinfo{volume}{B667}}, \bibinfo{pages}{1}
  (\bibinfo{year}{2008}).

\bibitem[{\citenamefont{Pumplin et~al.}(2002)}]{Pumplin:2002vw}
\bibinfo{author}{\bibfnamefont{J.}~\bibnamefont{Pumplin}} \bibnamefont{et~al.},
  \bibinfo{journal}{JHEP} \textbf{\bibinfo{volume}{07}}, \bibinfo{pages}{012}
  (\bibinfo{year}{2002}).

\bibitem[{\citenamefont{Dasgupta et~al.}(1998)\citenamefont{Dasgupta, Smye, and
  Webber}}]{Dasgupta:1998eqa}
\bibinfo{author}{\bibfnamefont{M.}~\bibnamefont{Dasgupta}},
  \bibinfo{author}{\bibfnamefont{G.~E.} \bibnamefont{Smye}}, \bibnamefont{and}
  \bibinfo{author}{\bibfnamefont{B.~R.} \bibnamefont{Webber}},
  \bibinfo{journal}{JHEP} \textbf{\bibinfo{volume}{04}}, \bibinfo{pages}{017}
  (\bibinfo{year}{1998}).

\bibitem[{\citenamefont{Sjostrand et~al.}(2008)\citenamefont{Sjostrand, Mrenna,
  and Skands}}]{Sjostrand:2007gs}
\bibinfo{author}{\bibfnamefont{T.}~\bibnamefont{Sjostrand}},
  \bibinfo{author}{\bibfnamefont{S.}~\bibnamefont{Mrenna}}, \bibnamefont{and}
  \bibinfo{author}{\bibfnamefont{P.~Z.} \bibnamefont{Skands}},
  \bibinfo{journal}{Comput. Phys. Commun.} \textbf{\bibinfo{volume}{178}},
  \bibinfo{pages}{852} (\bibinfo{year}{2008}).

\bibitem[{\citenamefont{Ellis et~al.}(2008)}]{Ellis:2007ib}
\bibinfo{author}{\bibfnamefont{S.~D.} \bibnamefont{Ellis}}
  \bibnamefont{et~al.}, \bibinfo{journal}{Prog. Part. Nucl. Phys.}
  \textbf{\bibinfo{volume}{60}}, \bibinfo{pages}{484} (\bibinfo{year}{2008}).

\bibitem[{\citenamefont{Vogt}(2001)}]{Vogt:2000ci}
\bibinfo{author}{\bibfnamefont{A.}~\bibnamefont{Vogt}}, \bibinfo{journal}{Phys.
  Lett.} \textbf{\bibinfo{volume}{B497}}, \bibinfo{pages}{228}
  (\bibinfo{year}{2001}).

\bibitem[{\citenamefont{Balazs and Yuan}(1997)}]{Balazs:1997xd}
\bibinfo{author}{\bibfnamefont{C.}~\bibnamefont{Balazs}} \bibnamefont{and}
  \bibinfo{author}{\bibfnamefont{C.-P.} \bibnamefont{Yuan}},
  \bibinfo{journal}{Phys. Rev.} \textbf{\bibinfo{volume}{D56}},
  \bibinfo{pages}{5558} (\bibinfo{year}{1997}).

\end{thebibliography}

%%%%%%%%%%%%%%%%%%%%%%%%%%%%%%%%%%%%%%%%%%%%%%%%%%%%%%%%%%%%%%%%%%%%%%%%%%%%%

\end{document}